\documentstyle[aps,prl,twocolumn]{revtex}
\begin{document}
\draft
\title{Quantum Fluctuations in Large-Spin Molecules}
\author{A.\ K.\ Zvezdin}
\address{Institute of General Physics, Moscow 117942, Russia}
\author{V.\ V.\ Dobrovitski and B.\ N.\ Harmon}
\address{Ames Laboratory, Iowa State University, Ames, Iowa 50011}
\author{M.\ I.\ Katsnelson}
\address{Institute of Metal Physics, Ekaterinburg 620219, Russia}
\date{\today}
\maketitle
\begin{abstract}
A new type of mesoscopic quantum effect in large-spin molecules
possessing easy-axis anisotropy, such as Mn$_{12}$, is predicted. 
The response of such a system to an external field applied 
perpendicular to the easy axis is considered. It is shown that
the susceptibility of this system exhibits a peculiar peak of purely
quantum origin. This effect arises from very general properties of
quantum fluctuations in spin systems. We 
demonstrate that 
the effect is entirely accessible for contemporary experimental 
techniques. Our studies show 
that the many-spin nature of the Mn$_{12}$ clusters is
important for a correct description of this quantum peak. 
\end{abstract}


Quantum fluctuations in spin systems receive much attention at present, 
being important in 
applied as well as fundamental fields of physics. 
As an example, mention can be 
made of attempts to explain the superconductivity in
high-T$_c$ cuprates in terms of electron interaction with 
quantum spin fluctuations \cite{scala}. 
Another example is the analysis of 
implementations of algorithms for quantum computations,
quantum cryptography and quantum searching \cite{quanta}, 
where quantum spin fluctuations enter into consideration. 
And, finally, quantum fluctuations are one of 
basic concepts underlying mesoscopic quantum effects in 
spin systems \cite {bcs}.

For experimental study of quantum fluctuations, 
molecular magnets, such as Mn$_{12}$, Fe$_8$,
Fe$_{17}$ etc. \cite{genclust}, are very suitable. 
They belong to a new class of magnetic 
materials that has been receiving increasing attention, being 
promising for fundamental studies as well as for 
practical applications. An adequate understanding of their 
properties is important. 

At present, the phenomenon of magnetization tunneling in
these magnets has attracted much attention \cite{jumps,barium}. 
But, as we show in this work, quantum fluctuations in 
molecular magnets can reveal themselves also in 
another, very interesting way, different from magnetization
tunneling and exhibiting a new 
type of mesoscopic quantum effect. 

We consider a mesoscopic magnetic system possessing an 
anisotropy of easy-axis type, such as the Mn$_{12}$ 
molecule, subjected to an external 
magnetic field directed perpendicular to the easy axis. 
We show that the susceptibility of this system as a 
function of the magnetic field has a clearly visible 
peak in the vicinity of the 
spin-reorientation transition. 
This peak is of purely quantum origin and 
differs significantly from what is expected for 
classical spins, thus 
demonstrating a fundamental difference between 
classical and quantum fluctuations in spin systems.

Our results demonstrate that the magnitude of the 
susceptibility peak 
noticeably depends on whether we account for the
internal structure of the cluster or use the single-spin 
description. This fact can be useful for elaboration 
of an adequate many-spin Hamiltonian for Mn$_{12}$.
It is an important issue now, since recent experiments 
\cite{excite} show that the single-spin model, long 
used for description of the Mn$_{12}$ cluster, is 
deficient and a model accounting for the constituent 
many-spin nature of these clusters is necessary.

To clarify the matter, let us consider 
a single-spin model (see inset in Fig.\ \ref{fig1}). 
Choosing the $z$-axis 
as the easy axis of the system and the $x$-axis
to coincide with the field direction, the Hamiltonian $\cal H$
of the system is 
\begin{equation}
\label{inham}
{\cal H}= -D{\cal S}_z^2-g\mu_{\text B}H_x {\cal S}_x,
\end{equation}
where $D$ is the easy-axis anisotropy constant, $g$ is the 
gyromagnetic ratio, $\mu_{\text B}$ --- the Bohr magneton,
$H_x$ is the field applied along the $x$-axis, 
${\cal S}_z$ and ${\cal S}_x$ are the spin operators. 
Here and below, the following numerical procedure has 
been used. 
The Hamiltonian of the system under consideration 
has been diagonalized numerically and the expectation value 
of the $x$-projection of the spin 
$\langle {\cal S}_x\rangle$ has been obtained by 
quantum-statistical 
averaging over the Gibbs canonical ensemble. The 
normalized susceptibility $\chi$ in the $x$-direction 
\begin{equation}
\chi(H_x)= d\langle {\cal S}_x\rangle / dH_x,
\end{equation}
has been calculated by numerical differentiation.
Since we 
are interested in quantum fluctuations only, we restrict ourselves
to the low-temperature region.

If we treat the system (\ref{inham}) as a classical spin at 
zero temperature, the dependence 
$\chi(H_x)$ has a step-like form 
(see Fig.\ \ref{fig1}, dashed line). But if we account for quantum 
fluctuations, i.e.\ if we consider the 
quantum spin described by the 
Hamiltonian (\ref{inham}), the peak in the magnetic susceptibility 
appears. 
In Fig.\ \ref{fig1} (solid line) we show the result for the case 
of the spin ${\cal S}=10$ with $g=2$ and the anisotropy constant 
$D=0.53$~cm$^{-1}$ (or, equivalently, $D/k_{\text B}=0.76$ K);
these parameters correspond to the single-spin model of Mn$_{12}$.

The essentially quantum nature of the effect can be demonstrated 
as follows. If ${\cal S}_x$ commuted with the Hamiltonian, 
as for classical spins, the 
susceptibility $\chi$ would be governed by classically defined 
fluctuations of ${\cal S}_x$:
$
\langle{\cal S}_x {\cal S}_x\rangle - \langle{\cal S}_x\rangle
  \langle {\cal S}_x\rangle.
$
Dependence of this quantity on $H_x$ (normalized by the value 
${\cal S}=10$) is presented in Fig.\ \ref{fig1} 
by the dotted line and shows a monotonic 
decrease with no peak.
But for a quantum spin the quantity ${\cal S}_x$ does not
commute with the Hamiltonian, and the correct measure of
fluctuations has to be formulated in terms of Kubo correlators:
$
\chi=({\cal S}_x,{\cal S}_x)=\int_0^{\beta} \langle 
\exp{({\cal H}\tau)}
{\cal S}_x \exp{(-{\cal H}\tau)} {\cal S}_x\rangle\,d\tau
-\beta\langle{\cal S}_x\rangle\langle{\cal S}_x\rangle, 
$
(where $\beta=1/k_{\text{B}}T$), which is different from the average 
due to non-commutativity of ${\cal S}_x$ and ${\cal H}$. The spectral 
densities of the Kubo correlator and the usual correlation function 
differ by the multiplier 
$(\beta\hbar\omega)/[1-\exp{(-\beta\hbar\omega)}]$
of purely quantum origin \cite{zubarev}.
We note that a sufficiently strong external field 
(comparable to the easy-axis anisotropy field) makes 
the relaxation time small enough to provide validity of 
the Kubo theory.

Now, for a more realistic description of the effect 
in Mn$_{12}$Ac molecules, we have to go beyond the single-spin model. 
The core of such a molecule, the cluster Mn$_{12}$,
schematically shown in Fig.\ \ref{fig2},
consists of 4 
Mn$^{4+}$ ions with spins 3/2 and 8 Mn$^{3+}$ ions with spins 2. 
The ions are coupled by exchange interactions; the 
values of the exchange integrals are not known, but 
the estimates are given in \cite{gat1}: 
$J_1=-150$ cm$^{-1}$ (AFM exchange),
$J_2=J_3=-60$ cm$^{-1}$ and 
$|J_4|<30$ cm$^{-1}$. These values are rough, but
describe correctly the scale of exchange 
interactions in Mn$_{12}$. In
the ground state the system has a large total spin, ${\cal S}=10$.
Recent experiments \cite{excite,hfepr} show that 
the excitations with spin values other than ${\cal S}=10$ are 
rather close to the ground state: the distance is 
40--60 K (exact values differ in different models), so 
an account of these excitations is necessary.
The cluster possesses rather strong 
magnetic easy-axis anisotropy: the zero-field splitting between
the levels $M=\pm 10$ and $M=\pm 9$ is 14.4 K.

The total number of spin states in Mn$_{12}$ is large even for 
modern computers, but we employ the 
fact that the exchange interactions $J_1$ are 
much larger than all 
the others, so corresponding pairs of Mn$^{3+}$ 
and Mn$^{4+}$ ions 
form dimers with total spin 1/2. This model has already been 
successfully used for a description of the 
spin states of the 
cluster \cite{gat1,zvezdin}. Its validity is supported by 
experiments \cite{bcs,gat1,hfepr,dotsenko}: 
the states of dimers with spin higher than 1/2 
(excitations of dimers) come 
into play when the external magnetic field 
is about 400 T or when temperature becomes as high as 
200--250 K. This scale of energies is 
completely irrelevant to the effect we are interested in: we 
consider magnetic fields of order of 7-10 T and temperatures 
of order of 1--4 K. 

Thus, we consider the Mn$_{12}$ cluster as
consisting of four "small" dimer spins 1/2 and four "large" spins 2
(corresponding to the four non-dimerized Mn$^{3+}$ ions),
coupled by exchange interactions. Moreover, we have to account
for anisotropic relativistic interactions in the cluster, so
the Hamiltonian of the system is:
\begin{equation}
\label{ham}
{\cal H}= -2J\sum_{\langle i,j\rangle} {\bf s}_i {\bf s}_j 
  -J'\sum_{\langle k,l\rangle} {\bf s}_k {\bf S}_l 
   + H_{\text{rel}},
\end{equation} 
where ${\bf s}_i$ are spin operators of small spins 1/2, 
${\bf S}_l$ are spin operators of large spins 2, and 
$H_{\text{rel}}$ denotes the relativistic Hamiltonian describing 
the magnetic anisotropy in the cluster. Summations in (\ref{ham})
are over pairs of spins coupled by exchange interactions.
In zeroth order of perturbation theory, $J=-J_2$ and $J'=-J_3+2 J_4$.
This Hamiltonian is the basis of the subsequent analysis.
Unfortunately, the exchange integrals of the Hamiltonian (\ref{ham})
and an exact form of the anisotropic term 
$H_{\text{rel}}$ are not known. Therefore, we examined 
the effect using very different sets of parameters. 
Below, we show typical cases, which give basic 
information about the dependence of the effect on the cluster 
parameters.

In Fig.\ \ref{fig3}, we show that the account of 
sufficiently large number of excited 
states is crucial for a correct description of the $\chi(H_x)$
dependence. If we include only 
the states belonging to the manifold ${\cal S}=10$, the height of 
the peak becomes considerably smaller. The more excited states 
we take into consideration, the more prominent the peak. 
Calculations with different sets of parameters 
give qualitatively the same results.

Positions of excited levels to a large extent are governed by
isotropic exchange interactions, which are not exactly known.
Therefore, 
we performed calculations with different values of exchange 
integrals, which yield the energies of excitations within the
region 30--60 K (cf. above). Some results are shown in 
Fig.\ \ref{fig4}; similar results have been obtained for 
different 
sets of exchange parameters and different forms of the 
anisotropic term $H_{\text{rel}}$ (see below).
Furthermore,  different types of 
easy-axis anisotropy in Mn$_{12}$ clusters have been checked.
Generally, the term $H_{\text{rel}}$
contains a single-site anisotropy of large spins 
(spins of Mn$^{3+}$ ions) and various kinds of 
anisotropic exchanges between different spins.
We performed calculations for three basic types of the 
easy-axis anisotropy:
\begin{mathletters}
\begin{eqnarray}
\label{a}
H^1_{\text{rel}}&=& -K\sum_{i=1}^4 \left(S_i^z\right)^2,\\
\label{b}
H^2_{\text{rel}}&=& -J_{zz} \sum_{\langle i,j\rangle} s_i^z
 s_j^z,\\
\label{c}
H^3_{\text{rel}}&=& -J_{Zz} \sum_{\langle i,j\rangle} s_i^z
 S_j^z,
\end{eqnarray}
\end{mathletters}
where summations in (\ref{b}) and (\ref{c}) are over 
exchange-coupled pairs of spins. Anisotropy parameters 
($K$, $J_{zz}$ or $J_{Zz}$) have been 
chosen to give a correct value of
the zero-field splitting between the states $M=\pm 10$ and
$M=\pm 9$ (14.4 K, see above).
All three types of anisotropy give rather 
close energies of low-lying excitations, and, correspondingly, 
the curves $\chi(H_x)$ are very close (the difference does not
exceed 6\%). 

Finally, we examined stability of the effect with respect 
to temperature variations and 
deviation of the external field direction. The results of 
our calculations for different 
temperatures are shown in Fig.\ \ref{fig5}. The peak shifts to 
lower field values and broadens. But,
until temperature becomes as high as 5 K, the peak is in evidence.
Also, as can be seen from Fig.\ \ref{fig6}, 
the direction of the external field 
should be close to the perpendicular to the easy axis. 
This restriction is not too stringent: the peak 
is visible until the 
deflection exceeds 2$^{\circ}$. These conclusions has been 
confirmed by calculations with different cluster parameters.

Our calculations show that the effect is
present in systems with very high value of the total spin ${\cal S}$.
E.g., for ${\cal S}=100$ the effect is in evidence 
at temperatures lower than
2--3 K, when the deflection of the field from the perpendicular to 
the easy axis is less than 0.2$^{\circ}$. Thus, this effect 
could be effective for study of mesoscopic quantum phenomena 
in systems with very large ${\cal S}$ (such as barium 
hexaferrite single particles \cite{barium}), where detection 
of resonant tunneling could be difficult.

Summarizing, we predict a new mesoscopic quantum effect.
The spin system possessing easy-axis anisotropy, e.g.\ 
the magnetic molecule Mn$_{12}$Ac, subjected 
to an external magnetic field directed perpendicular to 
the easy axis, exhibits a peak of susceptibility in the vicinity of 
spin reorientation transition. 
It is 
large and stable enough to be studied experimentally 
at temperatures about 2--4 K.
The external field magnitude should be 7--8 T, 
and possible deflection of the field direction from the 
perpendicular to the easy axis has to be not more than 1--2$^{\circ}$.
These conditions are quite accessible for contemporary 
experimental techniques.

Authors would like to thank M.-M. Teheranchi for his 
contribution at initial stage of the work and 
B.\ Barbara for helpful discussions. 
This work was partially carried out at the Ames Laboratory, which 
is operated for the U.\ S.\ Department of Energy by Iowa State 
University under Contract No.\ W-7405-82 and was supported by 
the Director for Energy Research, Office of Basic Energy Sciences 
of the U.\ S.\ Department of Energy. This work was partially supported 
by Russian Foundation for Basic Research, grants 96-02-16250 and 
98-02-16219.

\begin{figure}
\caption{The dependence of the magnetic susceptibility versus 
the external field (in Tesla): 
dashed line ---
classical system; solid line --- the quantum spin ${\cal S}=10$. 
Dotted line shows the classically defined fluctuations of
the quantity ${\cal S}_x$ normalized by the value ${\cal S}=10$.
The 
anisotropy constant is $D=0.53$~cm$^{-1}$ 
(or, equivalently, $D/k_{\text B}=0.76$ K). The inset shows the 
mutual arrangement of the applied field and the easy axis (EA).}
\label{fig1}
\end{figure}

\begin{figure}
\caption{Schematic plot of the Mn$_{12}$ cluster. Small black 
circles represent Mn$^{4+}$ ions, large white circles denote Mn$^{3+}$
ions. Different types of lines connecting the ions (solid, dashed,
dotted and dash-dotted) correspond to different types of exchange 
interactions ($J_1$, $J_2$, $J_3$ and $J_4$).}
\label{fig2}
\end{figure}

\begin{figure}
\caption{Dependence of the effect on the number of the excited levels 
included: (i) levels with energy up to 180 K 
are accounted for (solid line); (ii) levels with energy up to 90 K 
are included (dashed line); (iii) only the levels belonging 
to the $S=10$ manifold are accounted for (dotted line).
The easy-axis anisotropy is assumed to be 
caused by single-site anisotropy of "large" non-dimerized spins,
the form (4a). The cluster parameters are: $J'=90$ K, $J=0$,
$K=5.71$ K. Temperature $T=2$ K.}
\label{fig3}
\end{figure}

\begin{figure}
\caption{Dependence of the effect on exchange parameters:
(i) $J'=90$ K, $J=0$ (solid line); (ii) $J=-28$ K, $J'=140$ K
(dashed line); (iii) $J=-18$ K, $J'=90$ K (dotted line); 
(iv) $J=+18$ K, $J'=90$ K (dash-dotted line).
The easy-axis anisotropy is of 
single-site type, the form (4a), $K=5.71$ K. Temperature $T=2$ K.}
\label{fig4}
\end{figure}

\begin{figure}
\caption{Temperature dependence of the effect:
(i) temperature $T=2$ K (solid line); (ii) $T=3$ K (dashed line);
(iii) $T=4$ K (dotted line).
The easy-axis anisotropy is of single-site type,
the form (4a). The cluster parameters are: $J'=90$ K, $J=0$,
$K=5.71$ K.}
\label{fig5}
\end{figure}

\begin{figure}
\caption{Role of the deflection of the field from the normal 
to the easy axis: (i) no deflection (solid line);
(ii) the deflection is 1$^{\circ}$ (dashed line); 
(iii) the deflection is 2$^{\circ}$ (dotted line). 
The easy-axis anisotropy is of single-site type,
the form (4a). Cluster parameters are: $J'=90$ K, $J=0$,
$K=5.71$ K. Temperature $T=2$ K.}
\label{fig6}
\end{figure}

\end{document}